\documentclass[runningheads]{llncs}
\usepackage[utf8]{inputenc}
\usepackage[T1]{fontenc}

\usepackage{amsmath,amssymb}
\usepackage{bbm}
\usepackage{bm}
\usepackage{xspace}
\usepackage{standalone}
\usepackage{afterpage}
\usepackage[frozencache=true, cachedir=_minted]{minted}
\usepackage[kerning,spacing,tracking]{microtype}

\usepackage[x11names,table]{xcolor}
\usepackage[naturalnames=false]{hyperref}
\usepackage{graphicx}
\graphicspath{ {img/} }
\usepackage{tikz}
\usetikzlibrary{calc}
\usetikzlibrary{cd}
\usetikzlibrary{shapes}
\usepackage{csquotes}
\usepackage{booktabs}
\hypersetup{
  colorlinks = true,          
  urlcolor = blue, linkcolor = blue, citecolor = blue
}
\usepackage{svg}

\usepackage{csquotes}

\newcommand\polymake{\texttt{polymake}\xspace}

\newcommand\OSCAR{\texttt{OSCAR}\xspace}

\newcommand\Julia{\texttt{Julia}\xspace}
\newcommand\python{\texttt{Python}\xspace}
\newcommand\Python{\texttt{Python}\xspace}
\newcommand\JSON{\texttt{JSON}\xspace}
\newcommand\RNG{\texttt{RELAX NG}\xspace}
\newcommand\XML{\texttt{XML}\xspace}
\newcommand\OpenMath{\texttt{OpenMath}\xspace}
\newcommand\Mathematica{\texttt{Mathematica}\xspace}
\newcommand\Maple{\texttt{Maple}\xspace}

\newcommand\MIPLIB{\texttt{MIPLIB}\xspace}
\newcommand\Jupyter{\texttt{Jupyter}\xspace}

\newcommand\polyDB{\texttt{polyDB}\xspace}
\newcommand\MongoDB{\texttt{MongoDB}\xspace}
\newcommand\Nemo{\texttt{Nemo}\xspace}
\newcommand\Hecke{\texttt{Hecke}\xspace}
\newcommand\GAP{\texttt{GAP}\xspace}
\newcommand\Singular{\texttt{Singular}\xspace}
\newcommand\SageMath{\texttt{SageMath}\xspace}

\newcommand\Base{\texttt{Base64}\xspace}

\newcommand\PP{{\mathbb P}}
\newcommand\QQ{{\mathbb Q}}

\newcommand\ZZ{{\mathbb Z}}

\newcommand\GF[1]{\operatorname{GF}(#1)}

\title{A FAIR File Format for Mathematical Software}
\institute{Technische Universität Berlin, Chair of Discrete Mathematics/Geometry, Berlin, Germany
  \and  Max Planck Institute for Mathematics in the Sciences, Leipzig, Germany}

\author{Antony Della Vecchia\inst{1} \orcidID{0009-0008-1179-9862} \and \\ Michael Joswig\inst{1, 2} \orcidID{0000-0002-4974-9659} \and \\ Benjamin Lorenz\inst{1} \orcidID{0000-0003-2648-718X}}

\begin{document}

\maketitle

\begin{abstract}
We describe a \JSON based file format for storing and sharing results in computer algebra without losing accuracy.
Guided by practical usability, some key features are the flexibility to handle data structures unknown at the time of design,
a clear method for transitioning to the latest format and a way of separating data of distinct or even contradicting semantics.
This is implemented in the computer algebra system \OSCAR~\cite{OSCAR-book,OSCAR}, but we also indicate how it can be used in a different context.
\end{abstract}

We discuss general considerations, with a focus on comprehensibility and long-term storage.
General concepts for data serialization, like Protocol Buffers\footnote{\url{https://protobuf.dev/}} or \Julia's \cite{Julia-2017} \texttt{Serialization.jl}, do not suffice for the rich semantics of computer algebra.
Specialized software systems do allow for storing and writing files with mathematical data of a limited number of types, for example the \texttt{mps} file format used in optimization to
store linear and integer programs.
Hence this allows for sharing data, e.g., in databases such as \MIPLIB \cite{MIPLIB}.
However, formats like \texttt{mps} do not lend themselves to more general data.
The current standard for computer algebra systems is to use notebooks to store entire computations, \Jupyter\footnote{\url{https://jupyter.org}} being the current standard.
While these notebooks are very handy they do not provide a proper serialization of the intermediate results which can make certain recomputations undesirable or impossible.

In the late 1990s the \OpenMath project \cite{OpenMath:2000} developed a general framework for mathematical data.
Their effort was confronted with fundamental criticism, e.g., by Fateman \cite{Fateman:2001,Fateman:MVSD}. In light of the point held against
\OpenMath in \cite{Fateman:2001}, we pick a particular system, namely the new computer algebra system \OSCAR written in \Julia, and
rely on its semantics with no attempt to formalize the semantics in general.
We store data as annotated trees which is a common idea amongst comprehensive serialization formats, e.g., Protocol Buffers and \OpenMath.
Our format extends in a way the current \JSON file format of \polymake~\cite{polymake:2017}, which is a translation of the original \XML version \cite{polymake_XML:ICMS_2016}.
The syntax is fixed by an extensible \JSON schema, which is explained in Section~\ref{sec:schema}.
In Section~\ref{sec:beyond} we discuss how users of other computer algebra systems can make potential use of our format.

Our file format, developed as part of the Mathematics Research Data Initiative (MaRDI) \cite{the_mardi_consortium_2022_6552436}, aims to eventually extend beyond \OSCAR, for this reason we use the file extension \texttt{mrdi}.

\section{The File Format by Example}
\label{sec:example}

Our running example is a bivariate polynomial over a finite field, namely
\begin{equation}\label{eq:polynomial}
  2y^3z^4 + (a + 3)z^2 + 5ay + 1 \ \in \ \GF{49}[y, z] \enspace ,
\end{equation}
where $\GF{49}$, i.e., the finite field with 49 elements, is constructed as a degree two algebraic extension over the prime field $\GF{7}\cong\ZZ/7\ZZ$.
More precisely, as a $(\ZZ / {7 \ZZ})$-algebra, $\GF{49}$ is isomorphic to the quotient $(\ZZ / {7 \ZZ})[x]/ \langle x^2 + 1 \rangle$, and $x^2 + 1$ is a minimal polynomial.
In the latter quotient algebra we pick a generator and call it~$a$.
As the degree of the field extension equals two, the coefficients have a maximal $a$-degree of one.
\begin{figure}[bh]
   \begin{minted}[frame=single, framesep=3mm, linenos=false, fontsize=\footnotesize]{js}
{ "_ns": { "Oscar": ["https://github.com/oscar-system/Oscar.jl",
                     "1.0.0" ] },
  "_type": { "name": "MPolyRingElem",
             "params": "a7029443-b1d3-4708-a66d-f68eb6616fcf" },
  "data": [[["3", "4"], [["0", "2"]]],
           [["0", "2"], [["0", "3"], ["1", "1"]]],
           [["1", "0"], [["1", "5"]]],
           [["0", "0"], [["0", "1"]]]],
  "_refs": {
    "a7029443-b1d3-4708-a66d-f68eb6616fcf": { ... },
    "f2b7cb6b-535a-4a52-a0cc-75f8e93a6719": { ... },
    "23f25330-83f7-43a0-ac74-da6f2caa7eb8": {
      "_type": "FqField",
      "data": { "def_pol": {
        "_type": { "name": "PolyRingElem", 
                   "params": "f2b7cb6b-535a-4a52-a0cc-75f8e93a6719" },
        "data": [["0", "1"], ["2", "1"]]
} } } } }
\end{minted}
\caption{%
  \JSON description of the bivariate polynomial $2y^3z^4 + (a + 3)z^2 + 5ay + 1$ in the polynomial ring $\GF{49}[y, z]$ from \eqref{eq:polynomial}.
  We hide all but one reference describing a quotient field with defining polynomial $x^2+ 1$.
  \label{fig:json-polynomial}
}
\end{figure}

Our encoding keeps track of the entire history of the construction.
As a consequence, when we store that one polynomial, we also store the univariate polynomial ring $(\ZZ / {7 \ZZ})[x]$, the minimal polynomial $x^2 + 1$ and the quotient algebra $(\ZZ / {7 \ZZ})[x]/ \langle x^2 + 1 \rangle$.
In this way, we can associate with the algebraic expression \eqref{eq:polynomial} an annotated tree which reflects its construction.
This has a direct translation into \JSON code, shown in Figure~\ref{fig:json-polynomial}.

We distinguish between \emph{basic types}, which do have a fixed normal form.
These include standard \Julia types but also includes algebraic types such as the integers (\texttt{ZZRingElem}) or the rationals (\texttt{QQFieldElem}).
The more interesting types are \emph{parametric}, and the type \texttt{MPolyRingElem} of the polynomial \eqref{eq:polynomial} serves as our running example.
That type identifies \eqref{eq:polynomial} as an element in some multivariate polynomial ring.
Its ring of coefficients is referenced as a parameter to the type \texttt{MPolyRingElem}, where it listed under the \texttt{params} property.
The base ring of a multivariate polynomial ring can be any ring; in our example this is the quotient of a univariate polynomial ring by some ideal (spanned by one irreducible polynomial).
This gives rise to a recursive description because the parameter can have any type.
Consequently, parameters may have their own parameters.
All the parameter types, their parameters, and so on are stored in the global dictionary \texttt{\_refs}, and the \texttt{params} property refers to them via universally unique identifiers (UUIDs). We generate version four UUIDs specified by RFC 4122 on a first save and these persist throughout an active \OSCAR session.
For instance, this allows us to distinguish between isomorphic copies of a ring, which may play different roles in a specific computation.
This is useful, e.g., when we start with two polynomials $p\in\QQ[a,b]$ and $q\in\QQ[x,y]$, and much later we want to take their product in the ring $\QQ[a,b,x,y]$.
It occurs in daily computer algebra routine that the \enquote{universe} $\QQ[a,b,x,y]$ is not known in advance but rather the result of a sequence of several computational steps.

The example of finite fields illustrates that normal forms may not always provide an adequate description due to the incremental nature of certain computer algebra constructions, such as Hensel lifts in number theory; see \cite{BuchbergerLoos:1983} and \cite[\S15.4]{vonzurGathenGerhard:1999}. UUIDs facilitate tracking these evolving computations.

So, the type and its recursive parameters set the context which specifies the syntax of the serialized data.
The actual data is stored in the property with the same name.
In our running example the root of the data subtree has four children, one for each term of the polynomial \eqref{eq:polynomial}.
In the \JSON code each data subtree is written as a nested list of nested lists, marked by square brackets.

The discussion so far deals with the syntactic aspects of serialization.
It is a key design choice that the semantics is fully implicit.
In our example the semantics is determined by \OSCAR, version 1.0.0, which is specified in the namespace property \texttt{\_ns}.
Namespaces form the point of entry for the possibility to store data which are foreign to \OSCAR.
This is the topic of Section~\ref{sec:beyond} below.

\section{More Examples}
\label{sec:more}
To display the range of possibilities arising from our concept, we pick two examples of very different nature.

\paragraph{Non-general type surfaces in $\PP^4$.}
It is known that each smooth projective algebraic surface can be embedded in projective $5$-space, which we denote as $\PP^5$.
By a result of Ellingsrud and Peskine \cite{EllingsrudPeskine:1989} there are only finitely many families of surfaces of \emph{non-general type}, i.e., they admit an embedding already in $\PP^4$.
Decker and Schreyer obtained the number 52 as an explicit degree bound for such surfaces \cite{DeckerSchreyer:2000}.
In loc.\ cit.\ the authors also construct 49 non-general type surfaces in $\PP^4$ of degree up to 15.
That list is available in \OSCAR via files stored in our file format.

Here is one such surface (of degree three, with sectional genus zero), which is described as the vanishing locus of an ideal with three homogeneous generators in the polynomial ring $\GF{31991}[x,y,z,u,v]$.
The code below shows an interactive \Julia session using \OSCAR 1.0.0.
\begin{minted}[breaklines, fontsize=\footnotesize]{jlcon}
julia> S = cubic_scroll()
Projective scheme
  over finite field of characteristic 31991
defined by ideal with 3 generators

julia> defining_ideal(S)
Ideal generated by
  31990*x*y + 19122*x*z + 4788*x*u + ... + 20742*u*v + 25408*v^2
  7471*x*y + 23772*x*z + 27471*x*u + ... + 30545*u*v + 9903*v^2
  x^2 + 3601*x*y + 7253*x*z + 7206*x*u + ... + 6535*u*v + 26586*v^2
\end{minted}


Converting the descriptions of these 49 surfaces from the literature to objects suitable for computation takes some time and is prone to error.
So it is desirable to store such data explicitly, without the need of any computation or conversion.

\paragraph{Toric varieties.}
 The following code constructs two divisors on a toric variety; see~\cite{toric+varieties}.
 A \emph{toric variety} is an algebraic variety which is implicitly described by the normal fan of a convex lattice polytope,
 and a \emph{toric divisor} is a formal integer linear combination of facets of that polytope, i.e., rays of its normal fan.
 The polytope in our example is a triangle and so divisors are given by integer vectors of length three, one entry for each facet.
 
\begin{minted}{jl}
@testset "ToricDivisor" begin
  pp = projective_space(NormalToricVariety, 2)
  td0 = toric_divisor(pp, [1,1,2])
  td1 = toric_divisor(pp, [1,1,3])
  vtd = [td0, td1]
  test_save_load_roundtrip(path, vtd) do loaded
    @test coefficients(td0) == coefficients(loaded[1])
    @test coefficients(td1) == coefficients(loaded[2])
    @test toric_variety(loaded[1]) == toric_variety(loaded[2])
  end
end
\end{minted}
 
 The test saves and loads \texttt{vtd}, which is a vector formed of those two divisors, it then checks if loading yields the same objects.
 Additionally, the code checks if the underlying toric variety for the two divisors is the same, using UUIDs.

\section{Format Specification}
\label{sec:schema}

\JSON Schema~\cite{json-schema-2020-12} is a declarative language for describing \JSON file specifications, similar to \RNG for \XML.
Our file specification is shown in Figure~\ref{fig:file-schema}. 
\JSON has four types, namely \texttt{string}, \texttt{array}, \texttt{number}, and \texttt{object} (dictionary or hash map).
\begin{figure}[t]
\begin{minted}{js}
{ "$id": "https://oscar-system.org/schemas/mrdi.json",
  "$schema": "https://json-schema.org/draft/2020-12/schema",
  "type": "object",
  "required": ["_type"],
  "properties": { "_ns": { "type": "object" },
		  "_type": {
		    "oneOf": [
		      { "type": "string"},
		      { "type": "object", "properties": {
			"name": {"type": "string"},
			"params": {"$ref": "#/$defs/data"}
		      } } ] },
    "data": {"$ref": "#/$defs/data"},
    "_refs": {"type": "object", "patternProperties": {
      "^[0-9a-fA-F]{8}-([0-9a-fA-F]{4}-){3}[0-9a-fA-F]{12}$": {
	"$ref": "#" 
      } } } },
  "$defs": { "data": {
    "oneOf": [
      { "type": "string"},
      { "type": "array", "items": { "$ref": "#/$defs/data"} },
      { "type": "object", "not": { "required": [ "_ns" ] },
	"patternProperties": {
	  "^[a-zA-Z0-9_]*": {"$ref": "#/$defs/data"} } },
      { "$ref": "https://polymake.org/schemas/data.json"}
    ] } } }
\end{minted}
  \caption{%
    File format specification following the JSON Schema specification~\cite{json-schema-2020-12}.
    \label{fig:file-schema}
  }
\end{figure}
The first occurrence of the \texttt{type} property describes the file itself, where it expects the file to be of type \texttt{object}.
The \texttt{properties} and \texttt{patternProperties} keywords are used to describe the specifications for the keys and values of the object.
Only the values with keys being matched in the object specification (either exactly or by regular expression) will be checked.
The \texttt{required} keyword enforces that the objects have all properties listed in the array, here we enforce that the \texttt{\_type} property is present.
Some validators can handle common string formats, so we enforce that the keys of \texttt{\_refs} should have the format of a UUID.
The \texttt{oneOf} keyword is used to specify that one of the specifications in the list is expected.
The \texttt{\$ref} keyword uses a path or URL to refer to specifications defined elsewhere, the \texttt{\#} symbol denotes the root.
Other definitions can be described using the \texttt{\$defs} section. For example, our definition for \texttt{data} accepts several options
including recursive \texttt{object} and \texttt{array} structures as well as data formatted in accordance with \polymake's schema.

We use UUIDs instead of simpler indices so that references are valid throughout an entire session.
Consider the following scenario, Alice, computes with several, e.g., multivariate polynomials with coefficients in some fixed finite field, like in Figure~\ref{fig:json-polynomial}.
Then she stores a vector of three such polynomials in one file.
Further computations then yield a $3{\times}3$-matrix, whose coefficients lie in the same polynomial ring.
Alice stores that matrix in another file.
She sends both files to Bob, who wants to continue that computation, e.g., by multiplying the matrix with the vector.
In general, the finite field is constructed as a sequence of field extensions over the prime field.
While there is only one finite field of any given order, there are many field towers leading to the same.
Since the encoding depends on the details of the construction, the entire context must be present.
UUIDs allow for recognizing the same base ring across several files.
This is particularly useful for databases and large scale parallel computations.

\section{Beyond \OSCAR}
\label{sec:beyond}

The \texttt{mrdi} file format is meant to have a wide scope, and \OSCAR mainly serves as a proof of concept.
Here are some aspects not covered yet.
\paragraph{Namespaces.}
\OSCAR is based on and extends \Nemo/\Hecke \cite{nemo}, \GAP \cite{GAP4}, \polymake, and \Singular \cite{DGPS}.
So it is concerned with exact computations in number theory, group and representation theory, polyhedral geometry and optimization, as well as commutative algebra and algebraic geometry. 
Since our file format defers its semantics to a specific version of a specific software system, currently we are considering \enquote{algebraic} data only.
In \cite{Fateman:2001} Fateman pointed out that \enquote{Sin[x]} in \Mathematica\footnote{\url{https://www.wolfram.com/mathematica}} and \enquote{sin(x)} in \Maple\footnote{\url{https://maplesoft.com/products/maple}} mean very different things.
This makes it difficult to define and make use of any formal semantics covering such data beyond one software.
In our file format this problem could be resolved by defining separate namespaces for \Mathematica and \Maple.

It may even make sense to have data from distinct namespaces in the same file.
Any software system is free to interpret what it can understand and ignores the rest.
Via the underlying tree structure \enquote{the rest} may refer to arbitrary subtrees.
In this way, the \texttt{mrdi} file format is a flexible container format, which is similar in spirit to the Portable Document Format (PDF).\footnote{\url{https://pdfa.org/resource/iso-32000-pdf/}}
For instance, a PDF file may contain audio data, but not every PDF viewer is capable of playing back sound.

To show how this can work in practice, in Figure~\ref{fig:sage} we display a short code fragment which reads a multivariate polynomial with rational coefficients from a \texttt{mrdi} file into \SageMath \cite{william_stein_2020_4066866}.
That code is fully functional and complete, without shortcuts or hidden parts.

\begin{figure}[th]
\begin{minted}{py}
import json
from sage.all import PolynomialRing, QQ, prod

def load_oscar_polynomial(path):
  with open(path) as json_file:
    file_data = json.load(json_file)
    if ("Oscar" in file_data["_ns"] and
        file_data["_ns"]["Oscar"][1].startswith("1.0.")):
      t, d, refs = (file_data[k] for k in ["_type", "data", "_refs"])
      parent_ring_data = refs[t["params"]]["data"]
      base_ring = parent_ring_data["base_ring"]
      if base_ring["_type"] != "QQField":
        raise NotImplementedError("only rational coefficients supported")
      symbols = ",".join(parent_ring_data["symbols"])
      R, gens = QQ[symbols].objgens()
      p = R(0)
      for e, c in d:
        exps = [int(exponent) for exponent in e]
        coeff = QQ(c.replace("//", "/"))
        p += coeff * prod([g**i for g, i in zip(gens, exps)])
      return p
    else:
      raise RuntimeError("can only load OSCAR version 1.0 polynomials")
\end{minted}
  \caption{\Python code for \SageMath 10.2 to load a rational polynomial from a \texttt{mrdi} file written with \OSCAR 1.0}
  \label{fig:sage}
\end{figure}

Going through the \Python code also allows us to explain how we avoid Fateman's criticism of \OpenMath \cite{Fateman:2001,Fateman:MVSD}.
Code like the one in Figure~\ref{fig:sage} explicitly translates from \OSCAR to \SageMath.
This requires the programmer to know about both systems.
The necessity of direct communication between pairs of computer algebra systems was seen as a drawback in the 1990s, and it was a major motivation behind \OpenMath to overcome this obstacle.
The price to pay for a centralized communication concept like \OpenMath, however, is the need to formally specify the semantics.
Each computer algebra system has a rich implicit semantics which is often very difficult to spell out explicitly.
Consequently, developing a formalized semantics to govern several computer algebra systems simultaneously seems to be at least as involved as writing an entirely new computer algebra system from scratch.
It is therefore our conclusion to stick to the implicit semantics of one system, e.g., \OSCAR, and to translate explicitly whenever necessary.
In this way Fateman's criticism does not apply.

For rational polynomials the effort to translate from \OSCAR to another computer algebra system is quite moderate, as illustrated in Figure~\ref{fig:sage}.
Depending on the data types, other translations might require more effort.
Namespaces create the flexibility for every user to pick their own point of departure, i.e., picking a computer algebra system other than \OSCAR.

\paragraph{Databases.}
Any serialization lends itself to storing similar files in some systematic folder hierarchy, mimicking a simple database.
Our file format is no exception, and the algebraic surfaces from Section~\ref{sec:example} form an example in \OSCAR.
The No-SQL database \MongoDB uses a record structure which essentially agrees with \JSON objects.
In this way, our serialized data can directly be used for storage and retrieval in highly efficient large scale databases.
The same concept was already exploited in \polymake's database project \polyDB \cite{polydb}. Note that \MongoDB requires
UTF-8 encoded strings in the JSON objects, whence we restrict our file format to UTF-8, too.

\begin{figure}
  \centering
  \includegraphics[height=0.27\textwidth]{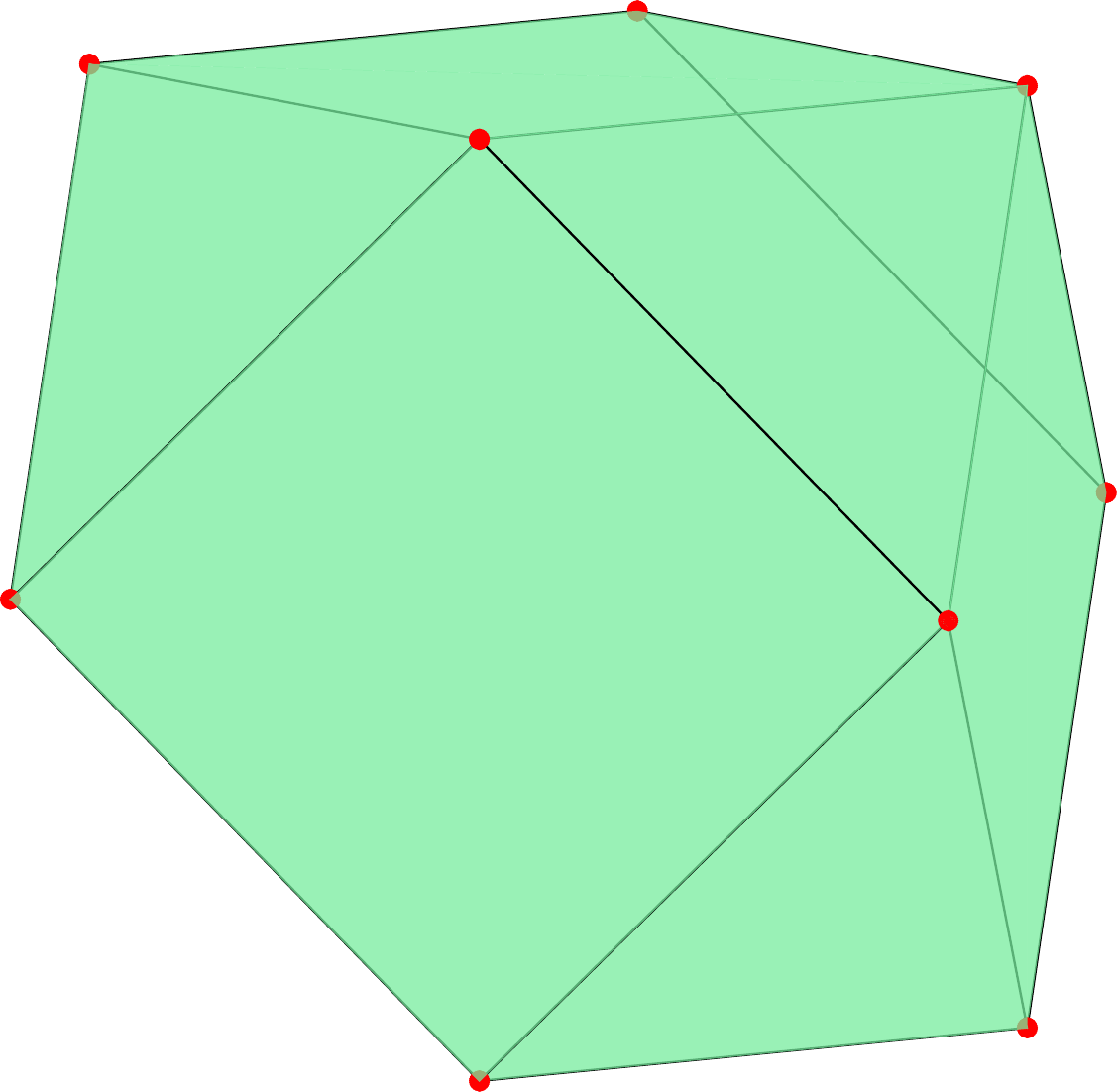}\qquad
  \includegraphics[height=0.27\textwidth]{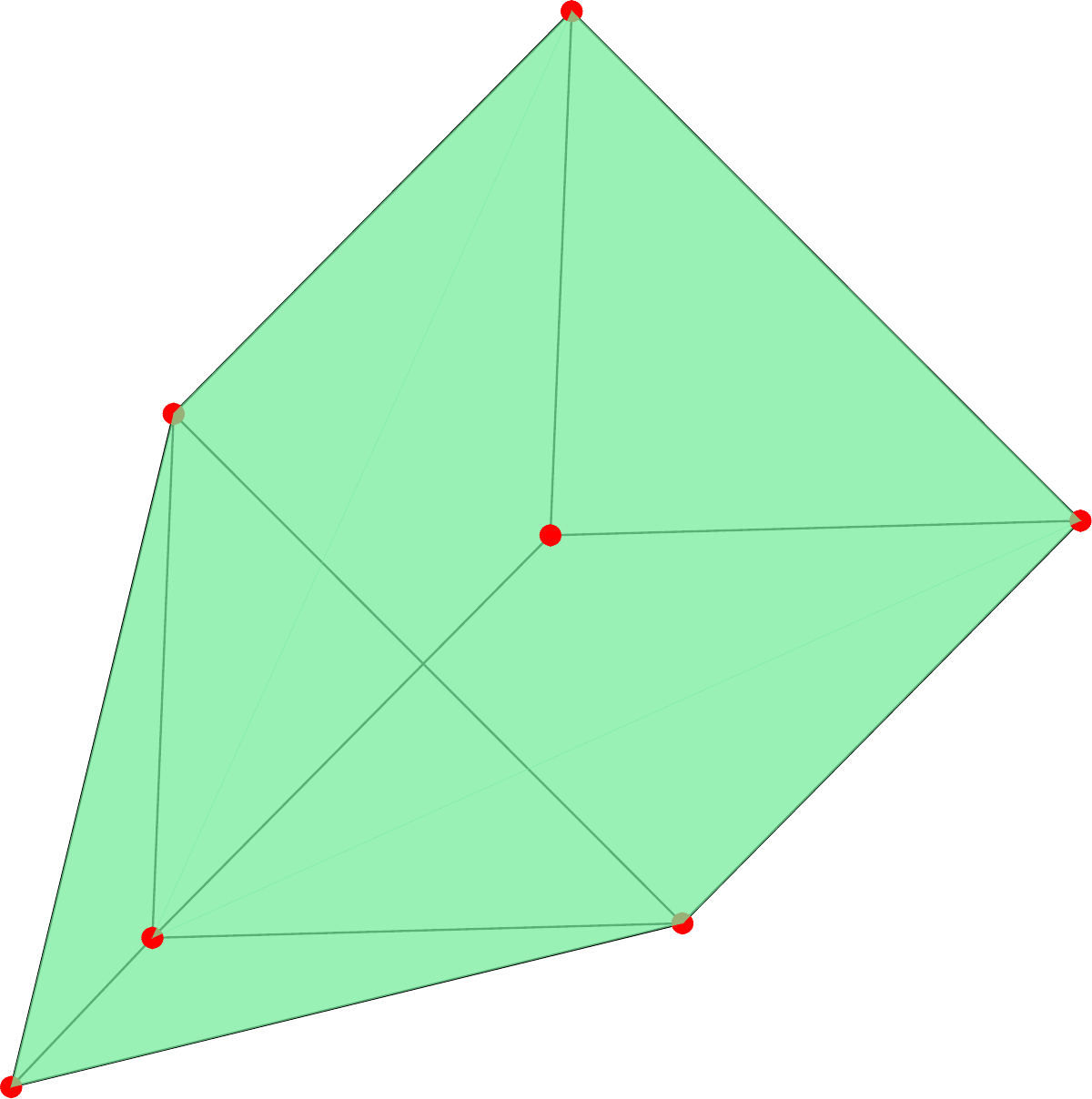}\qquad
  \includegraphics[height=0.27\textwidth]{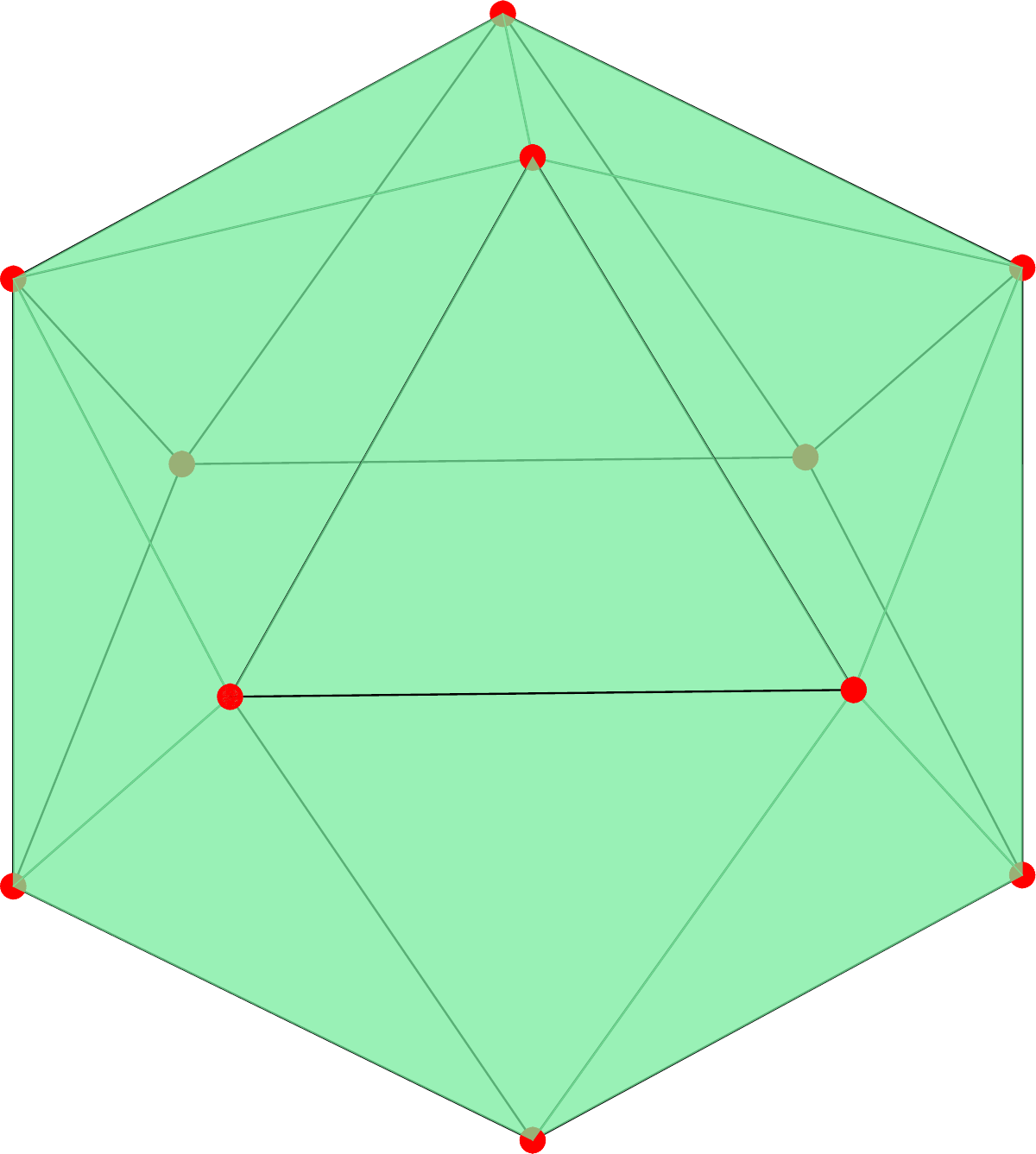}
  \caption{(a) Triangular cupola, (b) elongated triangular pyramid, (c) gyroelongated pentagonal pyramid}
  \label{fig:johnson}
\end{figure}

Another interesting collection of \texttt{mrdi} files are the Johnson solids from \cite{geiselmann_2024_10729583}.
The latter are 3-dimensional convex polytopes whose facets are regular polygons.
The Johnson solids generalize the Archimedean solids, and there are precisely 92 of them (up to rigid motions and scaling) which are not Archimedean.
The dataset \cite{geiselmann_2024_10729583} comprises exact algebraic numbers as well as approximate floating point numbers as coordinates for those 92 polytopes.
Furthermore, the data comes with a \Julia script allowing users to read certain properties of a Johnson solid using only standard \Julia and \JSON parsing, independent from \OSCAR.
Since the \texttt{mrdi} file format is based on \JSON, there are even simpler methods to access basic information about that dataset.
For instance, via \texttt{jq},\footnote{\url{https://jqlang.github.io/jq/}} which is a command-line \JSON processor.
The following three commands print out the respective number of vertices for the triangular cupola, the elongated triangular pyramid and the gyroelongated pentagonal pyramid.
These three Johnson solids are displayed in Figure~\ref{fig:johnson}.
\begin{minted}[breaklines, fontsize=\footnotesize]{jlcon}
> jq '.data.float.VERTICES | length' j3
9
> jq '.data.float.VERTICES | length' j7
7
> jq '.data.float.VERTICES | length' j11
11
\end{minted}

\paragraph{Version upgrades.}
The \OSCAR project itself should not be considered monolithic or complete.
On the contrary, \OSCAR has been designed to keep evolving, with new data types, encodings and use cases.
Some of these changes suggest modifications to how the data is serialized, and the file format needs to be able to go along with such changes.
To this end our data comes with version numbers.
Upgrade scripts provide arbitrary transformations from old data into the current standard, in contrast to Protocol Buffers whose format
is forward compatible.
Such an upgrade scheme is in place within the \polymake project for more than a decade.
In 2020 version 4.0 of \polymake replaced the previous \XML based serialization by \JSON, through the same mechanism.
This shows that such upgrades are feasible.

\paragraph{Metadata.}
The current version of the file format can optionally attach metadata that includes an entry for the name of the data and an author ORCID.\footnote{\url{https://orcid.org}} We think this is sufficient for a first version of the file format and is subject to change depending on the requirements set by the MaRDI portal.\footnote{\url{https://portal.mardi4nfdi.de/wiki/Portal}} The MaRDI portal aims to provide services for the findability and accessibility of mathematical research data.

\section{Concluding Remarks}
The main point of our design is the lack of dependence on any particular programming language. This by itself sets it apart from \Julia's \texttt{Serialization.jl} or \python's \texttt{pickle} module.
Converting mathematical data into \JSON objects, which are mere strings, always means an overhead.
For long term storage space efficiency is often less relevant, while other features are more important.
However, once data becomes so large that it hits physical bounds, e.g., as the capacity of a hard drive, it becomes mandatory to think about data compression.
Via namespaces our approach allows to compress subtrees of the \JSON object and use \Base binary to text encoding to obtain a new valid \JSON object.
A more thorough discussion is beyond the scope of the present article.

\subsection*{Acknowledgements}
We are grateful to the entire \OSCAR developer team for implementing and discussing code; special thanks to Claus Fieker, Tommy Hofmann, and Max Horn.
Further we are indebted to Lars Kastner for discussing FAIR principles, to Wolfram Decker for explaining algebraic surfaces in $\PP^4$, and to John Abbott,
Ewgenij Gawrilow, and Aaruni Kaushik for helpful feedback.


\bibliography{biblio}

\end{document}